\documentclass{aa}  
\usepackage{graphicx}
\usepackage{natbib}
\usepackage{txfonts}
\begin{document}
   \title{Line and continuum variability of two intermediate-redshift, high-luminosity quasars}

   \author{D.Trevese \inst{1},
           D.Paris \inst{1},
           G. M. Stirpe \inst{2},
	   F.Vagnetti \inst{3},
          \and
	  V. Zitelli \inst{2}
          }

   \institute{Dipartimento di Fisica, Universit\`a di Roma ``La Sapienza'', P.le A.Moro 2, 
   I-00185 Roma(Italy)\\
              \email{dario.trevese@roma1.infn.it}
         \and
             INAF - Osservatorio Astronomico di Bologna, via Ranzani, 1 - 40127 Bologna (Italy) \\
	  \and 
	  Dipartimento di Fisica, Universit\`a di Roma ``Tor Vergata'', 
	  Via della Ricerca Scientifica, 1, I-00133 Roma (Italy)
             }

   \date{}

 
  \abstract
   {It has been shown that the luminosity of active galactic nuclei and the size of their broad line region
obey a simple relation of the type $R_{BLR}=a L^{\gamma}$, from faint Seyfert nuclei to bright quasars,
allowing  single-epoch determination of the central black hole mass $M_{BH}= b L^{\gamma} \Delta^2_{H_{\beta}}$
from their luminosity $L$ and width of  $H_{\beta}$ emission line. Adopting this mass determination for cosmological
studies requires the extrapolation to high redshift and luminosity of a relation whose calibration,
relies so far on reverberation mapping measurements performed for $ L \la$10$^{46}$erg s$^{-1}$ 
and  redshift  $z \la$ 0.4.}
{We initiated a campaign for the spectrophotometric  monitoring of a few luminous, intermediate redshift
    quasars whose apparent magnitude, $V<15.7$, allows observations with a 1.8 m telescope, aimed at
proving that emission lines  vary and  respond to continuum variations even for luminosities $\ga$10$^{47}$erg s$^{-1}$,
and  determining eventually their $M_{BH}$ from reverberation mapping.}
{We have repeatedly performed simultaneous spectrophotometric observations of quasars and reference stars
to determine relative variability of continuum and emission lines. We describe the observations and methods of analysis.}
   {For the quasars  PG 1634+706 and PG 1247+268  we obtain light-curves respectively for CIII]($\lambda\lambda$1909\AA), 
   MgII($\lambda\lambda$2798\AA) and for CIV($\lambda\lambda$1549\AA), CIII]($\lambda\lambda$1909\AA) emission lines 
 with  the relevant continua.
During 3.2 years of observation, in the former case no continuum variability has been detected and the evidence for
line variability is marginal, while in the latter case both continuum and  line variability are detected with 
high significance and the line variations appear correlated with continuum variations.}
   {The detection of the emission line variability in  a quasar with  $L \sim  $10$^{47}$erg s$^{-1}$  
encourages the prosecution of the monitoring campaign which should provide a black hole mass estimate
in other 5-6 years, constraining the mass-luminosity relation in a poorly explored range of luminosity.}


\authorrunning{D. Trevese et al.}
\titlerunning{Line and continuum variability}

   \maketitle
%

\section{Introduction}

Supermassive black holes  (SMBHs) are believed to inhabit  most, if not
all, the bulges of  present-epoch galaxies \citep{korm95}, 
and  strong  evidences exist of  a correlation
between  the  black hole  mass  and  either  the mass $M_{bulge}$  and  luminosity
\citep[][and refs. therein]{marc03} or the  velocity dispersion $\sigma_*$ of the
host  bulge \citep{ferr00,trem02}.  This strongly
suggests  that  the formation  and  growth  of  SMBHs and galaxies  are
physically  related processes and provides a basis for a theory of  cosmic 
structure formation,  incorporating the  feedback from Active Galactic Nuclei (AGNs) 
\citep[][and refs.therein]{silk98,vitt05}.
Black hole masses determinations based on stellar or gas kinematics are intrinsically
limited, by angular resolution, to relatively nearby objects and cannot be applied to 
bright AGNs where the nuclear light prevails over the galactic component, just in the 
central region where the galactic gas or star motion is dominated by the black hole 
gravitational field.
The reverberation-mapping  technique does not suffer of this 
limitation and represents the only mean to measure the mass of SMBH in bright AGNs.

Emission lines, in the optical-UV region, are interpreted as recombination of a gas which is
photoionized by the continuum radiation emitted by the inner region of the nucleus, 
presumably an accretion disk surrounding the black hole.
Emission lines respond to variation of the ionizing continuum. 
Although the  physical origin of these variations is poorly known \citep[e.g.][]{trev02,vand04,devr05}
it is possible to use the response of lines to continuum variations to investigate the 
structure of the line emitting region. This requires long campaigns of accurate spectrophotomeric
monitoring of AGNs, which have led in the past to major progresses towards  understanding the physics of 
the ``atmosphere'' of Seyfert 1  galaxies. A summary of these results
is given in \citet{pete93}.

Line widths, e.g. the FWHM $\Delta_{H_{\beta}}$  of the $H_{\beta}$ emission line,
correspond to r.m.s. velocities of the emitting gas clouds.
A cross-correlation analysis
of continuum and emission-line light-curves, evidencing a time delay $\tau$ of line respect
to continuum variations, allows to estimate the size $R=\tau/c$  of the 
region where the line photons are generated. If  the gas motion in the Broad Line Region (BLR)  
is dominated by gravitation \citep{pete00}, the size estimate $R_{BLR}$ can be combined with the line
width to yield a {\it primary} estimate of the virial mass of the black 
hole $M_{BH} \propto \Delta_{H_{\beta}}^2/G R_{BLR}$
and the relevant Eddington ratio. For Seyfert 1 galaxies typical BLR sizes are of the order 
of light-days to light-weeks. Similar studies are more difficult
for quasars (QSOs) which require a longer monitoring. 

A long term campaign for a subsample of 28 QSOs 
was started in 1991 with the Wise 1.0m  and the Steward 2.3m telescopes \citep{maoz94}.
As a result, nine years later \citet{kasp00}  provided mass estimates for the entire sample.
The new data, combined with previous results on Seyfert 1 galaxies, thus spanning a much wider 
range of intrinsic luminosity, allowed to establish  an average relation between the intrinsic luminosity 
and the size,  $R_{BLR}=a L^{\gamma}$, with $\gamma \simeq 0.7$,  which allows a {\it secondary} estimate of the black hole
mass based on single-epoch observations of luminosity and line width: 
$M_{BH}=b L^{\gamma} \Delta_{H_{\beta}}^2$, where both the constant $b$ and $\gamma$ are determined statistically
from the available echo-mapping data. 
Recent studies show that $\gamma$ is in the range 0.5-0.7, depending on how luminosity is defined, which lines
are selected for the echo-mapping and the fitting procedure adopted \citep{kasp05,bent06}.  

The  extreme importance of secondary mass estimates relies on the fact
that on the basis of single epoch observations it
is possible to study the evolution in cosmic time of the mass distribution of QSOs/AGNs,
and to extend the studies of the relation existing between QSOs and host bulges properties.
However, the above correlations with primary masses, based on echo-mapping,
were established for relatively close and faint AGNs with $z \leq 0.4$ and 
[$\lambda L_{\lambda}(5100 \AA)\la 10^{46}$ erg s$^{-1}$], thus it is 
presently unknown whether
they can be extrapolated to higher luminosities and/or redshifts. For instance, the extrapolation
of the $M_{BH}-L$ relation \citep{kasp00}, together with the assumption that the known $M_{BH}-M_{bulge}-\sigma_*$
relations holds \citep{trem02} , leads to predict the existence of galaxies with 
$M_{bulge} \sim 10^{13.1}-10^{13.4} M_{\odot}$ and $\sigma_*$ exceeding 800 km s$^{-1}$. Such galaxies 
 have never been observed, and  their existence would put important constraints on galaxy formation models \citep{netz03}.
Therefore it is essential to extend  the primary mass measures to higher redshifts and luminosities.
On the other hand, for high QSO luminosities a large size of the broad line region is expected.
This would cause both a smoothing of the line light-curve and larger time delays with respect to continuum variations \citep{wilh05},
thus the very detectability and the amplitude of line variations are open questions.

A sample of objects with redshifts in the range $2<z<3.4$ and apparent magnitude as faint as $m_V \sim $18
is being monitored by \citet{kasp04}  with the 9m Hobby-Eberly Telescope \citep[HET;][]{rams98}
and new results have been presented recently \citep{kasp06}. During their 6-year monitoring of 6 QSOs, significant continuum and
emission-line variations were detected in all targets and a preliminary black hole mass estimation is given for one of them.

In the present paper we describe a new monitoring campaign limited to objects with $V<15.7$ and  $1 < z < 4$
which, thanks to their apparent brightness, can be observed with the medium-small  1.82 m Copernicus Telescope at Cima Ekar
(Asiago, Italy), 
through a service mode scheduling of a long  term monitoring,  and  allow to investigate whether:
i) echo-mapping is feasible for objects as bright as  $\lambda L_{\lambda}(5100\AA) \sim 10^{47}$ erg  s$^{-1}$
and ii) the  $R_{BLR}$-luminosity correlation can be extrapolated to such brightness. 
The paper is organized as follows. Section 2 describes the sample, observations and the data reduction procedure.
Section 3 describes the results for two quasars of the sample. Section 4 summarizes the results and 
discuss  future prospects.
In the following we derive $\lambda L_{\lambda}(5100\AA)$  from the flux in the Johnson V band, extrapolating
the flux density to the rest-frame $\lambda=5100 \AA$ with a power law $f_{\nu} \sim \nu^{-0.5}$, and  
assuming a standard cosmology  $H_o=70$ km s$^{-1}$ Mpc$^{-1}$, $\Omega_M=0.3$,  and $\Omega_{\Lambda}=0.7$.

\section{Observations}

\subsection{Object selection}

The sample has been extracted from the \citet{vero03} (11th ed.) catalog with the condition $\delta >0$,
$V< 15.7$ mag and $z>1$ in order to select  objects of bright enough intrinsic luminosity
to investigate the bright end extension of the 
$R_{BLR}$ vs. $\lambda L_{\lambda}(5100\AA)$ relation \citep{kasp00,kasp05}.
These conditions identify 12 objects,  only four of which were monitored, owing to the 
limits on observing time. These four objects are listed in Table 1.

\begin{table}
\caption{The quasars monitored}
\label{table:1}
\centering               
\begin{tabular}[h]{lccc}
\hline\hline             
Object  & $z$  & $V$ & $\log [\lambda L_\lambda(5100{\rm\AA})]$\\
&&&[erg s$^{-1}$]\\
\hline
APM 08279+5255  &  3.911 & 15.20   & 47.7\\
PG 1247+268         &  2.042  & 15.60   & 47.0\\   
PG 1634+706         &  1.337  & 15.27   & 46.7\\ 
HS 2154+2228       &  1.290  & 15.30   & 46.7\\
\hline
\end{tabular}
\end{table}

\begin{figure}
   \centering
\resizebox{\hsize}{!}{\includegraphics{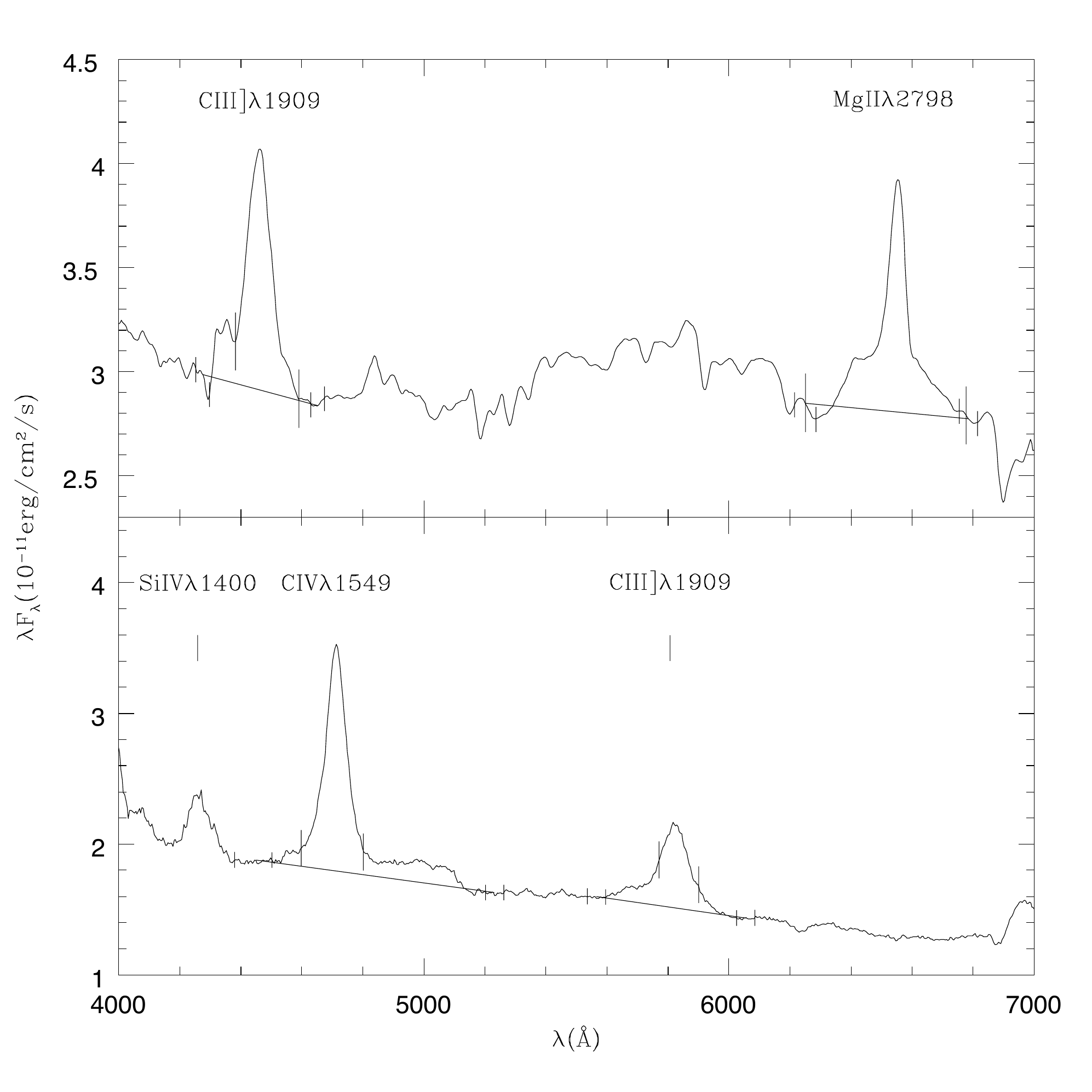}}

      \caption{
Average spectra of PG 1634+706 (upper panel) and PG 1247+268 (lower panel).
Spectral ranges for continuum determination (short ticks) and line flux evaluation (long ticks), 
as reported in Table 2, are shown.
}
         \label{Fig1}
   \end{figure}

Observations at intermediate redshift  allow to sample the variability
of MgII  $\lambda2798$, CIII] $\lambda1909$,  CIV $\lambda1559$ lines,
instead of $H_{\alpha}$, $H_{\beta}$, $H_{\gamma}$ observed in the low
redshift  study of  \citet{kasp00}.  This allows  to  study either  BLR
at smaller  distance from the center, or  regions of the same sizes
of those producing the Balmer lines, but using lines which respond to different
part of the continuum spectrum. The  main  emission  lines  falling  within  the  observed
wavelength interval  are indicated for  two objects in Table  2.
Figure \ref{Fig1} shows the average spectra of PG 1634+706  and PG 1247+268.


With respect to  other QSO monitoring  programmes, ours is the  first which
includes the MgII  line in part of the observed  sources. This line is
particularly  significant,  because   i)  past  Seyfert  1  monitoring
campaigns conducted with IUE have shown  its lag to be similar to that
of $H_\beta$ \citep{clav91,reic94}; ii) this line is
most  often  used  to   derive  estimates  of  black-hole  mass  from
single-epoch  spectra of high-$z$  QSOs \citep{mclu02},  because  
its  width  is tightly  correlated  to  that  of $H_\beta$.
  Deriving a lag  for the  MgII line  would therefore  allow to
estimate  the black-hole mass  most consistently  with respect  to the
results  of Seyferts,  which are  mostly  based on  the monitoring  of
$H_\beta$.

APM 08279+5255 is one of the most luminous known QSOs if its emission is considered isotropic.
However it has been shown to be lensed by a foreground galaxy \citep{irwi98}. Three components,
separated by a few tenths of an arcsec have been detected in  near-infrared images obtained with Keck telescope
and different models of the lensing field predict
a few days delay between photometric variations of the components \citep{ledo98,ibat99,egam00}: a short time compared
with the expected time scale of intrinsic variations. APM 08279+5255 is also a Broad Absorption Line QSO \citep{irwi98},
which makes more difficult to define regions free from either emission or absorption features
to measure continuum variations.  The analysis of this object is deferred to a forthcoming paper.
HS 2154+2228 has been observed so far only 5 times and the analysis requires further monitoring. 
The other two objects PG 1247+268 and PG1634+706 are analyzed in this paper with the aim of :
i) verifying the adequacy of the observational data and reduction procedures, under the assumption that variability amplitudes 
and characteristic time scales can be extrapolated from the properties of
fainter objects; ii) possibly detecting line variations in objects as luminous as 
$\lambda L_{\lambda}(5100\AA)\sim  10^{47}$ erg  s$^{-1}$, and compare their amplitude with continuum variations.
For both objects a star of comparable magnitude, as close as possible to the QSO, has been selected
for the relative spectrophotometric calibration, described in the next section.

\subsection{Spectrophotometric Observations and Data Reduction}

Observations were carried out at  Asiago 1.82 m telescope
equipped with the Faint Object Spectrograph \& Camera  AFOSC  
which is a focal reducer with  reduction factor of 0.58, designed to allow
a quick switching between spectroscopic and imaging modes. The  scale at the  focal plane is  21.7 $^{\prime\prime}$/mm.
The detector is a 1024x1024 thinned CCD
array TEK1024  with  22x22 $\mu$m$^2$ pixels corresponding to 
a scale of 0.473 arcsec pixel$^{-1}$ and a FOV of 8.14x8.14 arcmin$^2$.
We adopted a  8".44-wide slit  and a  grism with  a 
dispersion  of 4.99  \AA\ pixel$^{-1}$, providing a typical resolution
of $\sim 15$ \AA\ in the spectral  range 3500-8450 \AA. 
Spectrophotometric exposures are performed after orienting the slit to include both the QSO and the
reference star of comparable magnitude, located at (12:50:11.5 +26:33:32) and (16:34:57.4  +70:32:49) (J2000)
for PG 1247+268 and  PG 1647+706  respectively. The reference stars are included as internal calibrators
for QSO spectra, as described by \citet{maoz90} and \citet{netz90}.
The wide slit is necessary to avoid different fractional losses of the QSO and star light
due to possible non perfect slit alignment and differential refraction, which could cause
spurious variation of the flux ratios.
Lamp flats and Hg-Cd ark spectra were also taken for wavelength calibration.
At each epoch, typical science observations  consist  of two  consecutive
exposures of 1800 s.

QSO and star spectra,$Q(\lambda)$ and $S(\lambda)$  were extracted with the standard IRAF procedures.
The QSO/star ratio 
 as a function of
wavelength  is  computed  for   each  exposure $k=1,2$
\begin{equation}
\mu^{(k)}(\lambda)=Q^{(k)}(\lambda)/S^{(k)}(\lambda).
\end{equation}

\noindent
This  quantity is
independent  of extinction  changes during  the night.  This  allows a
check of consistency between the two exposures and the rejection of the data
if inconsistencies occur (under the assumption that QSO variations are negligible on ~1 h time scale).
In fact, typical spectra of two consecutive exposures have a ratio  
$\mu^{(1)}(\lambda)/\mu^{(2)}(\lambda)$
of order unity, with  deviations  
smaller than 0.02 when averaged over ~500 \AA, at least in the
$4000\AA-7000\AA$ range. When discrepancies are larger than 0.04 both  exposures are rejected.

It is important to notice that, because of the large slit width, there are small
changes in the $\lambda$ scale due to changes of the object position within the slit
(which are in general negligible in the case of pairs of consecutive exposures).
This suggested to proceed as follows. 
As a first step we register the zero point of the $\lambda$ scale. However, since the changes of position within the slit
do not correspond exactly to  rigid shifts of the  $\lambda$ scale, we select portions of the spectra in wavelength intervals
$|\Delta \lambda| \la 1000 \AA$ around each of the QSO emission lines considered (see Table 2),
and we determine the shifts for each portion.

Once the wavelength scales are registered, 
the QSO and star  spectra taken in the two exposures
are co-added and the ratio 
\begin{equation}
\mu_i(\lambda)=(Q_i^{(1)}+Q_i^{(2)})/(S_i^{(1)}+S_i^{(2)})
\end{equation}
is computed, at each epoch  $i$.
The reference star is flux calibrated  at a reference epoch.
Since our aim is to compute relative flux variations, we reduce all quasar spectra to this reference epoch, 
by multiplying all the $\mu_i(\lambda)$
by the  flux calibrated star spectrum $f^S(\lambda)$:  $f^Q_i(\lambda) \equiv  \mu_i(\lambda) f^S(\lambda)$.
We stress that we are not interested in the absolute flux calibration, whose accuracy is of the order of 20\% and which is applied
for the sole purpose reporting the QSO spectra in physical units. The star spectrum  $f^S(\lambda)$ adopted is the same for all epochs,
thus it does not affect {\it relative} flux changes we want to measure.

The QSO spectrum $f^{Q}(\lambda)=c(\lambda)+l(\lambda)$ can be 
decomposed in a line $l(\lambda)$ and continuum  $c(\lambda)$ spectra.
Two values of the continuum,  $c_{short}$ and $c_{long}$ at  shorter and longer wavelength with respect to
the most prominent QSO emission lines,
are evaluated in regions as free as possible of other emission features, defined by the wavelengths ranges
($\lambda_{short,1}$,$\lambda_{short,2}$) and ($\lambda_{long,1}$,$\lambda_{long,2}$) respectively.
Table 2 reports the observer-frame wavelengths  defining the continuum regions and the intervals for line integration.
These were chosen on the basis of the analysis of \citet{clav91,reic94} of UV spectra of low redshift AGNs observed
with the International Ultraviolet Explorer (IUE), with slight modifications to maximize the S/N ratio 
($c_{long}$ of MgII($\lambda 2798$ \AA) falls outside the wavelength range covered by the IUE spectrograph).

\begin{table*}
\caption{Wavelength intervals for lines and continua [\AA]}
\label{table:2}
\centering       
\begin{tabular}{lllllllll}
\hline\hline             
Object&$z$&line&$\lambda_{short,1}$&$\lambda_{short,2}$&$\lambda_1$&$\lambda_2$&$\lambda_{long,1}$&$\lambda_{long,2}$\\
\hline
PG1634+706 &1.337     &      $CIII] \lambda1909$ &$4252$     &$4298$     &$4382$  &$4590$ &$4630$     &$4674$     \\
PG1634+706 &1.337     &      $MgII \lambda2798$  &$6215$     &$6285$     &$6251$  &$6777$ &$6755$     &$6815$     \\
PG1247+268 &2.042     &      $CIV \lambda1549$   &$4380$     &$4502$     &$4598$  &$4801$ &$5202$     &$5262$     \\
PG1247+268 &2.042     &      $CIII] \lambda1909$ &$5535$     &$5595$     &$5770$  &$5900$ &$6025$     &$6085$     \\

\hline
\end{tabular}
\end{table*}

Line fluxes are computed as:
\begin{equation} 
f_l = \int_{\lambda_1}^{\lambda_2}[F^{(Q)}(\lambda) - c^{int}(\lambda)] d \lambda,
\end{equation} 
where $c^{int}(\lambda)$ is the linear interpolation through $c_{short}$ and $ c_{long}$.
The extremes $\lambda_1$ and ${\lambda_2}$, also listed in Table 2, not necessarily coincide with
 $\lambda_{short} \equiv (\lambda_{short,1}+\lambda_{short,2})/2 $ and  
$\lambda_{long} \equiv (\lambda_{long,1}+\lambda_{long,2})/2$, and are chosen to optimize the $f_l$ signal to noise ratio.

 Direct B  and R images in 15x15  arcmin fields centered on  the QSOs were
 taken  to check  possible variability  of the  reference star. 
Comparison of the reference star with the brightest star in the field showed a r.m.s. fractional flux variation of ~0.02 
in both cases. Thus we can assume that the reference star is not variable, at this level of accuracy.

\begin{figure}
   \centering
\resizebox{\hsize}{!}{\includegraphics{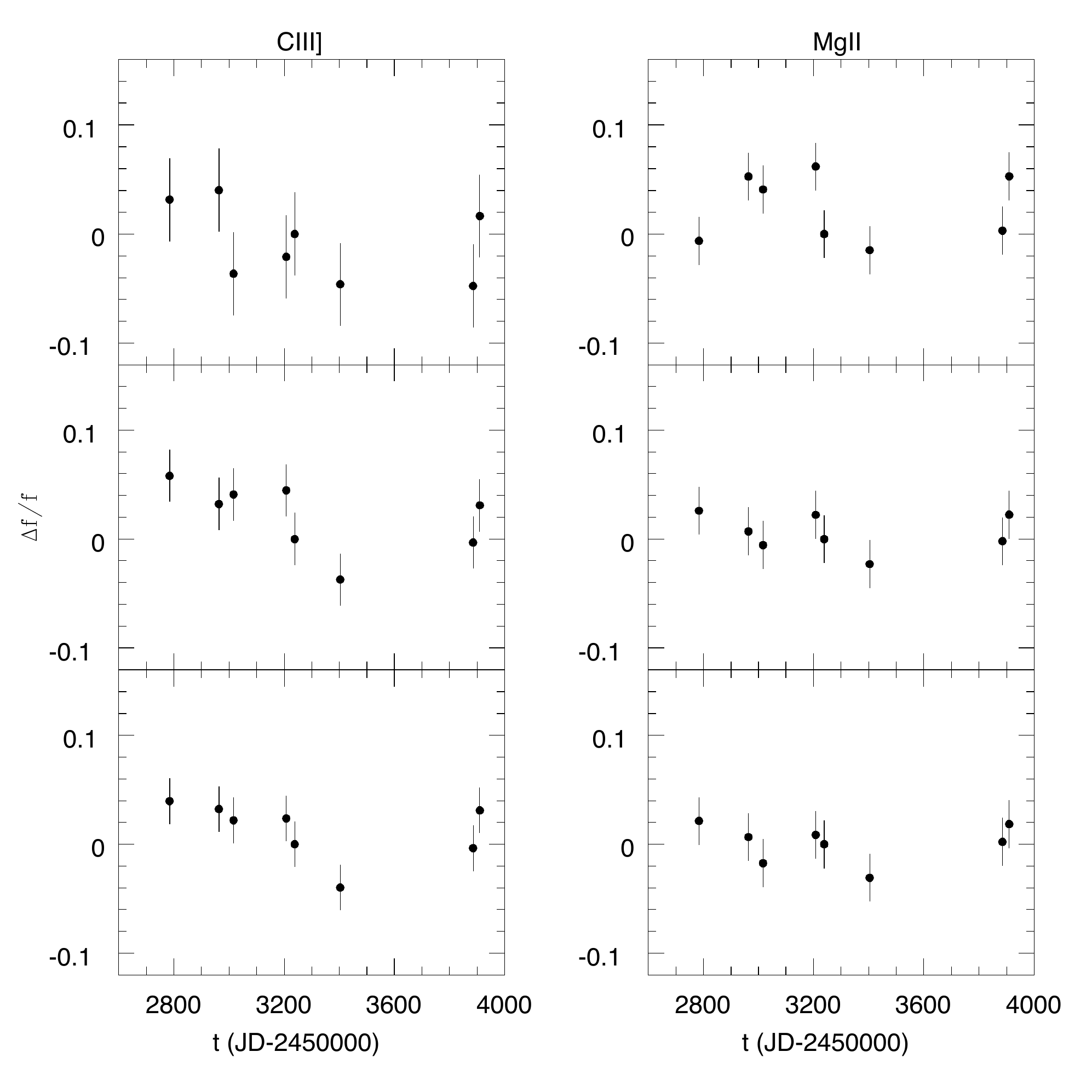}}

      \caption{
Light-curves of PG 1634+706 in the observer's frame, as relative flux variations $\Delta f/f$. 
Upper panels: emission lines, middle panel: shorter wavelength continuum  $c_{short}$,
lower panel: longer wavelength continuum $c_{long}$, left:  CIII]/$\lambda$1909 \AA,
right: MgII $\lambda$2798 \AA.
}
         \label{Fig2}
   \end{figure}

\begin{figure}
   \centering
\resizebox{\hsize}{!}{\includegraphics{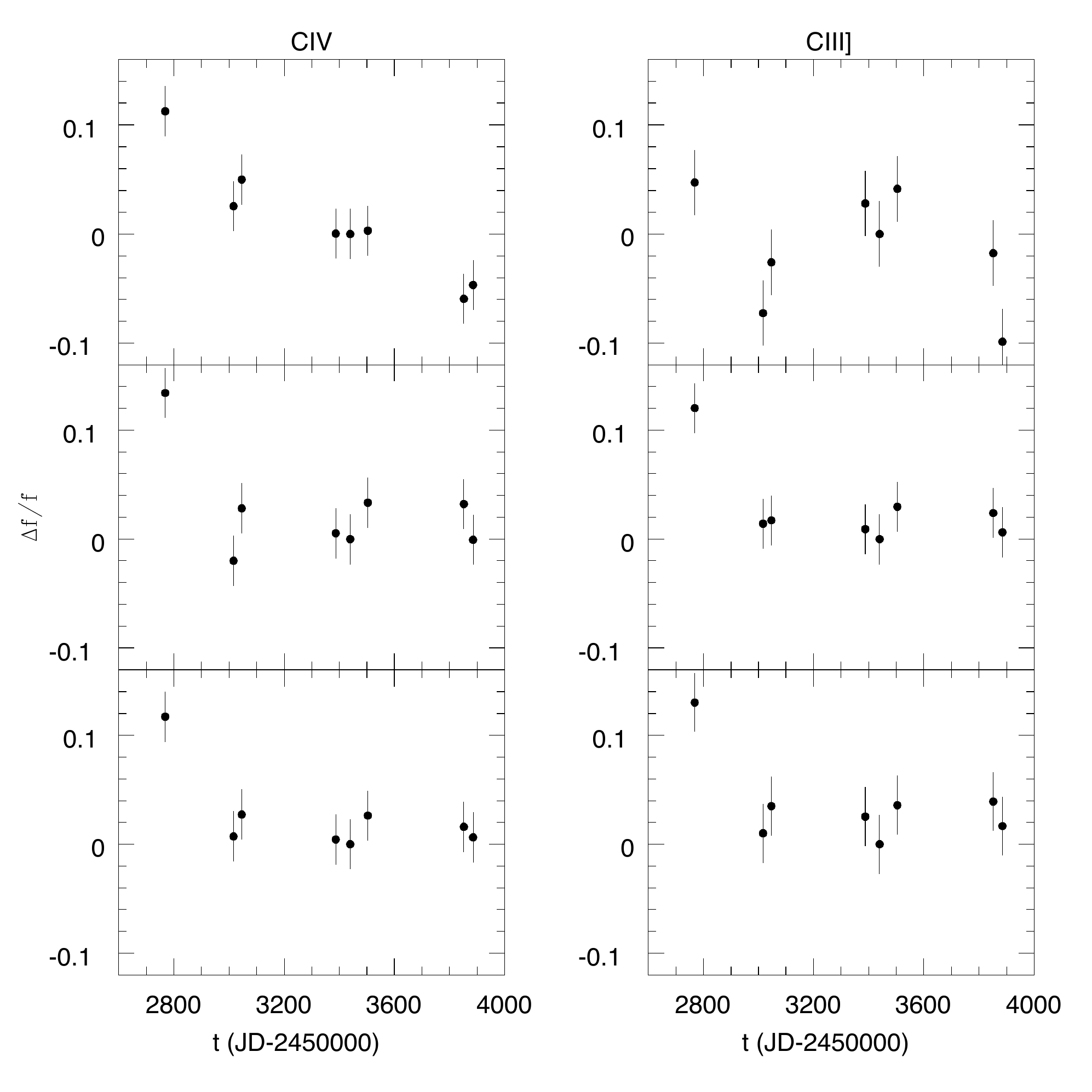}}

      \caption{
Light-curves of PG 1247+268 in the observer's frame, as relative flux variations $\Delta f/f$. 
Upper panels: emission lines, middle panel: shorter wavelength continuum  $c_{short}$,
lower panel: longer wavelength continuum $c_{long}$, left:  CIV/$\lambda$1549 \AA,
right: CIII] $\lambda$1909 \AA.
}
         \label{Fig3}
   \end{figure}

Relative-flux calibration 1-$\sigma$ errors, reported in figures \ref{Fig2} and \ref{Fig3}, are estimated 
by adding in quadrature the r.m.s. uncertainty on the flux of the reference star  
to the r.m.s. fractional uncertainties of line or  continuum fluxes, which, in turn, are computed by comparing
pairs of exposures which are added to form quasar spectra of individual  epochs.


\begin{figure}
   \centering
\resizebox{\hsize}{!}{\includegraphics{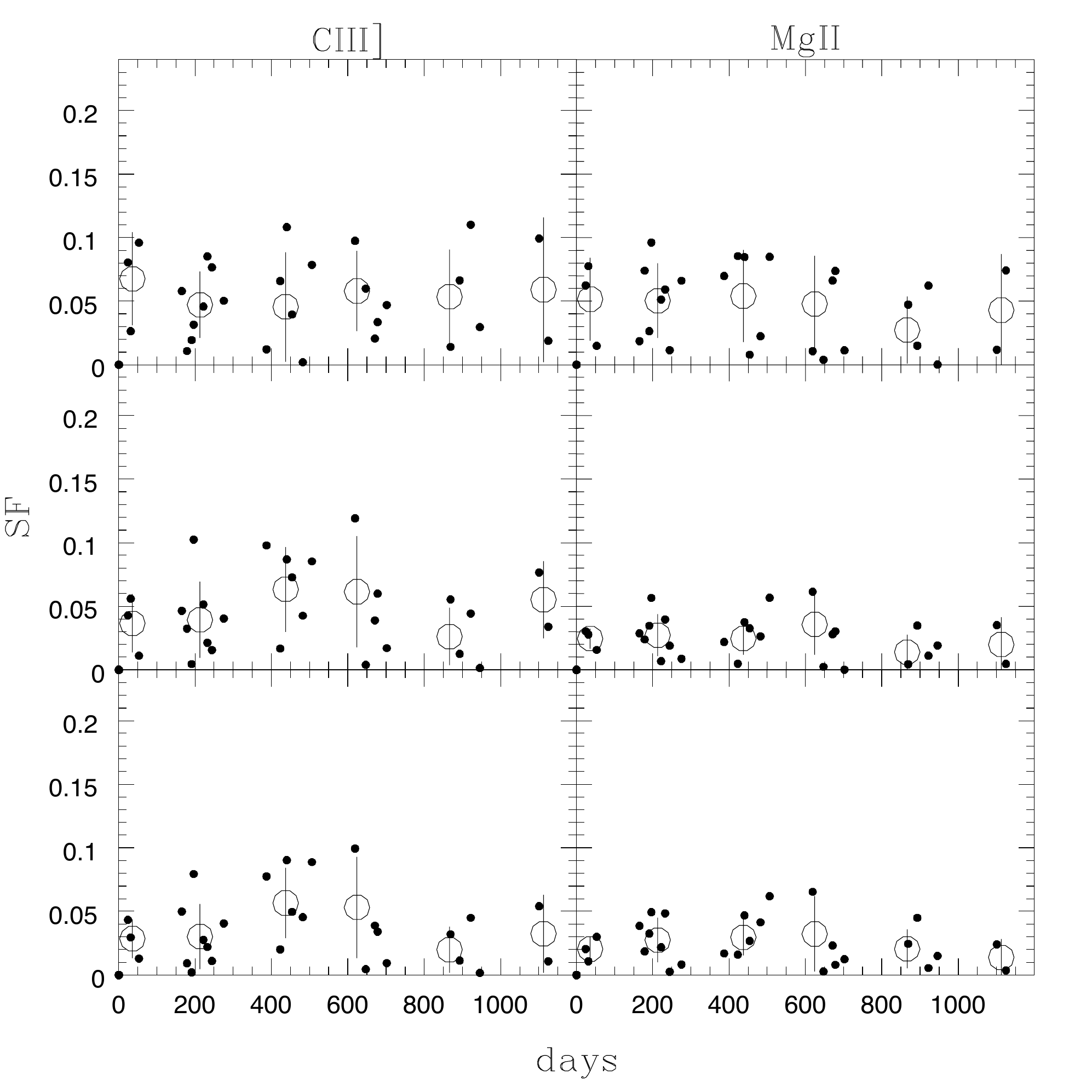}}

      \caption{
Structure functions for PG 1634+706 in the observer's frame. 
Upper panels: emission lines, middle panel: shorter wavelength continuum  $c_{short}$,
lower panel: longer wavelength continuum $c_{long}$, left:  CIII]/$\lambda$1909 \AA,
right: MgII $\lambda$2798 \AA.
}
         \label{Fig4}
   \end{figure}


\begin{figure}
   \centering
\resizebox{\hsize}{!}{\includegraphics{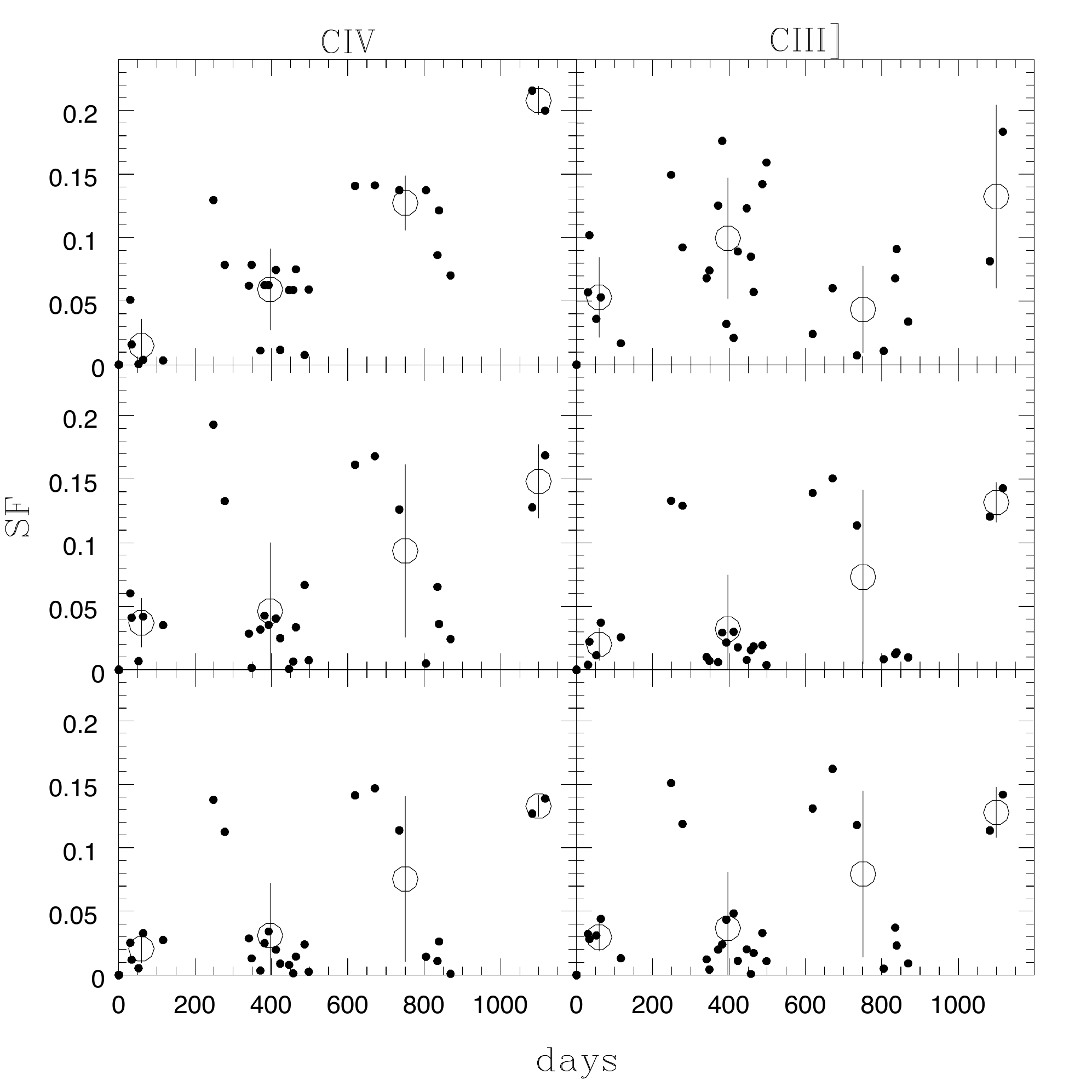}}

      \caption{
Structure functions for PG 1247+268.Upper panels: emission lines, middle panel: shorter wavelength continuum  $c_{short}$,
lower panel: longer wavelength continuum $c_{long}$, left:  CIV/$\lambda$1549 \AA,
right: CIII] $\lambda$1909 \AA.
}
         \label{Fig5}
   \end{figure}

\section{Results and discussion} 

Figures \ref{Fig2} and  \ref{Fig3}  show the light-curves in the observer's frame,  
for emission lines and continua listed in Table 2, for PG1634+706
and PG1247+268 respectively.
The light curves are expressed in terms of the relative flux variations
$\Delta f/f$, with respect to the flux $f$ at the reference epoch
where the absolute calibration was performed (see Section 2.2). 
The total time base is 3.3  years in the observer's frame, corresponding in the rest-frame to 1.4 yr  for
the former QSOs and 1.1 yr for the latter.
In the case of PG1634+706  no significant variations are detected: 
neither in CIII] line, which has the lowest S/N ratio due to its intrinsic faintness,
nor in the MgII line. The relevant continua are also consistent with no variability at a level of 
$\Delta f/f \sim$ 0.02-0.03 r.m.s.. 

The result is different in the case of PG 1247+268. The faintest line CIII] $\lambda$1909 \AA\ shows a marginal evidence
of variability and the stronger line CIV $\lambda$1549 \AA\  seems to decrease steadily during the observing period.

An almost steady decrease does not allow to derive possible line-continuum time delays
from cross-correlation analysis.
However we can establish quantitatively the evidence of continuum and, most important, of line variability.
For this purpose
we define the unbinned discrete structure function 
\begin{equation}
UDSF(\tau_{ij})= \sqrt{\frac {\pi}{2}}|y(t_i)-y(t_j)|, 
\end{equation}
\noindent
where $y(t)$ represents 
any of the line or continuum light-curves considered,  
$t_i$ and $t_j$ are two observation epochs and $\tau_{ij}=t_i-t_j$ is the time delay. 
The (binned) structure function can be defined in bins of time delay,  centered at $\tau$: 
\begin{equation}
S(\tau)=\frac{1}{M}\left[\sum_{i,j} UDSF(\tau_{ij})\right], 
\end{equation}
\noindent
where the sum is extended to all the $M$ values of UDSF belonging to a given bin of  $\tau$.
We adopt the average of UDSF, instead of UDSF$^2$,  since it is more stable, i.e. less sensitive to deviant points; 
the factor $\sqrt {\frac {\pi}{2}}$ in equation (4) makes $S(\tau)$ equal to the standard deviation in the case
of a Gaussian distribution \citep[see][]{dicl96}. 
Structure functions for lines and continua of the two quasars PG 1634+706 and PG 1247+268 are reported in Figures 
\ref{Fig4} and \ref{Fig5} respectively, as a function of the time delay in the observer's frame.
The error bars reported in the figures represent simply the standard deviation of the UDSF in each bin and not the uncertainty on their 
average value. The latter would be $1/(M-1)$ times the standard deviation, for $M$ uncorrelated  UDSF values.

In the case of PG 1634+706 the structure function analysis simply confirms what already appears from the light-curves,
namely  no significant variations of lines and continua are detected. The r.m.s. noise level is about 0.02-0.03 for continua and 0.05 for lines.

On the contrary, in the case of PG 1247+268, there is a clear increase of variability for long time lags.
To establish the significance of this variability it is necessary to evaluate 
the probability of the null hypothesis that a 
value of $S(\tau)$ is produced by pure noise.
For this purpose we generated  mock noise light-curves ,  $n(t_i)$, i=1,$N_{epo}$, 
where $N_{epo}$ is the number of observing epochs, by
extracting random numbers with a Gaussian distribution of standard deviation $\sigma$,
which represents pure noise,
under the assumption that the photometric noise is not correlated on the time scale of our minimum sampling interval (namely about 10 days).
Then, the relevant structure function $S_{mock}(\tau)$ has been computed 
with the same binning adopted for real data. The simulation was iterated for $10^7$ times to generate, for each bin, the statistical distribution of the  $S_{mock}(\tau)$ values. 

The value adopted for $\sigma$ has been estimated, conservatively, from the observed 
structure functions of figure  \ref{Fig4} and \ref{Fig5}
themselves, as the value of $S(\tau)$ in the first bin. 
This is, in fact, an overestimate of noise, since it includes short-time-scale variations
of the QSO, which however should be negligible on the basis of previous studies of (fainter) QSOs \citep{give99,kasp00}. 
The almost constant value
of the structure functions in the case of PG 1634+706 confirms 
the adopted hypothesis that noise is not correlated on long time scale.

For  the CIII] line the r.m.s. noise  ($\sim 0.05$) is larger than in the case of the CIV line ($\sim 0.02$).
This depends on the intrinsic faintness of the CIII] line.
According to the above simulations  the probability of the null hypothesis, for the second and fourth bins,
is $P(>S) \sim 4 \times 10^{-4}$.
The local maximum of  $S(\tau)$, for CIII] at $\tau \sim$400 days,
 is due to an ``oscillation'' of the light-curve on a time scale comparable
with the total time base of the observations. This indicates that a longer time base is needed to
properly average individual oscillations and to obtain a ``stationary'' structure function.
Further sampling, trying to avoid yearly periodicity due to periods of optimal observability, 
is needed to obtain the resolution required for time delay measurements. 

The structure function of CIV line shows a much more significant variability: 
the probability of the null hypothesis in the case of the last two bins, where $S(\tau)$ is 0.13 and 0.21 respectively,
is less than $10^{-6}$.
The systematic decrease of CIV line luminosity and the relevant continua might seem to rely on the 
value of the light-curves at the first epoch. Thus we have recomputed the $S(\tau)$ after removing the first epoch.
Indeed, in the case of  continua the systematic decrease of $S(\tau)$ is no longer evident. 
On the contrary, for the line it remains practically unchanged
(except that the last bin  is no longer present). 
Thus the fact that the probability of the null hypothesis is smaller than $10^{-6}$ does not rely 
on that particular point in the light-curve.

The  main goal of  new echo-mapping  campaigns is  to obtain  a direct
measure of  the central black hole  mass and to  establish whether the
scaling of  QSO broad line region  size and masses remain  the same at
higher quasar luminosity.  The very detectability of line variation in
the most  luminous QSOs was,  until recently, an open  question, since
3C273 with ($\lambda L_{\lambda}(5100  \AA) \la 10^{46}$ erg s$^{-1}$)
was the  brightest QSO with detected  line variability \citep{kasp05}.
An evidence of line variability  in brighter objects has been reported
more  recently  by  \citet{kasp06}.   The present  detection  of  line
variability  in one  of  the two  quasars  considered strengthens  the
notion that quasar emission lines respond to continuum changes in high
luminosity  QSOs like  in  lower luminosity  ones  and indicates  that
echo-mapping campaigns are feasible for $\lambda L_{\lambda}(5100 \AA)
\ga 10^{47}$ erg  s$^{-1}$.  

Our results are consistent  with those of
\citet{kasp06},  who  have  monitored  sources  of  comparable
luminosities  and redshifts.  We  have also  detected variability,  of
comparable  amplitude, in  the continuum  and CIII]  and  CIV emission
lines in one of our  sources. Considering the shorter time baseline of
our observations, the  fact that variability has not  been found in PG
1634+706 is consistent with  the segments without variations which can
be seen in the \citet{kasp06} light curves on this time scale.

To estimate the further monitoring needed to measure line-continuum delays, we should take into account that,
on average, the amplitude of variability increases with the rest-frame time lag.
However this dependency is weak: 
$\log S(\tau_{rest}) \sim 0.15 \log \tau_{rest}$,
according to previous statistical studies on large quasar samples \citep{vand04,devr05}.
In the case of our quasars  PG 1634+706 and PG 1247+268, the observer's frame
$\tau_{obs}=\tau_{rest}(1+z)$ is dilated by  a factor 2.3 and 3 respectively. The size of the broad line region,
based on Balmer line reverberation-mapping,
is expected to scale with luminosity as $R_{BLR} \propto L^{\gamma}$ with $\gamma$=0.5-0.7, depending on how $L$ is measured
and the $R_{BLR}$ vs. $L$  fit is obtained
\citep{kasp00,kasp05,bent06,kasp06}.
A Balmer line BLR of about 3 light years in the rest frame is expected for the relevant luminosities.
On the other the  hand CIV line corresponds to a BLR size 2 times smaller than Balmer lines \citep{pete00}. 
Considering that in general a time baseline of about twice the light-crossing time is required in order to obtain a 
reliable lag from the light curves, we estimate that reverberation mapping should be feasible after 5-6 more years of monitoring.

\section{Summary}
\begin{itemize}
\item We have initiated a campaign for the monitoring of 4 high luminosity QSOs and we present the result
for two of them on the basis of 3.3 years of observation with the 1.82 m telescope of the Asiago  Observatory.
\item We discuss the data reduction, the procedures adopted and the level of accuracy attained in 
 relative spectrophotometric variability measurements.
\item We perform a structure function analysis of the light-curve and numerical simulations
to establish the confidence level of line variability detection.
\item We detect line  and continuum variability in one of the two QSOs, PG 1247+268 
of $\lambda L_{\lambda}(5100 \AA)= 10^{47}$ erg s$^{-1}$, with a probability
of less than 10$^{-6}$ of the null hypothesis that the observed structure function is produced by pure noise. 
\item This detection supports the notion that emission lines respond to continuum variations as in 
substantially less luminous QSOs.
\end{itemize} 
The results encourage the prosecution of the campaign which should provide time delay, and black hole mass, estimates
in 5-6 years.

\begin{acknowledgements}
We acknowledge  support from the Asiago Observatory team, in particular  Hripsime Navasardian,
for observations.
\end{acknowledgements}

\end{document}